\documentclass[aps,pra,epsfig,showpacs,superscriptaddress,footnote]{revtex4}
\usepackage{amsmath,amsthm,amssymb}
\usepackage{graphics}
\makeatletter
\def\be{\begin{equation}}
\def\ee{\end{equation}}
\def\bea{\begin{eqnarray}}
\def\eea{\end{eqnarray}}
\def\beaa{\begin{eqnarray*}}
\def\eeaa{\end{eqnarray*}}
\def\bma{\begin{mathletters}}
\def\ema{\end{mathletters}}
\def\bi{\begin{itemize}}
\def\ei{\end{itemize}}

\newcommand{\ket}[1]{ | \, #1  \rangle}

\begin{document}

\title{Quantum Cournot equilibrium for the Hotelling-Smithies model of product choice}

\author{Ramij Rahaman}
\email{Ramij.Rahaman@ii.uib.no} \affiliation{Department of Informatics, University of Bergen, Post Box-7803, 5020, Bergen,
Norway }
\author{Priyadarshi Majumdar }\email{majumdar_priyadarshi@yahoo.com}
\affiliation{Jyotinagar Bidyasree Niketan H.S. School, Kolkata 700
108, India}
\author{B. Basu}
\email{banasri@isical.ac.in}\affiliation{Physics and Applied
Mathematics Unit, Indian Statistical Institute,
Kolkata-700108,India}

\begin{abstract}
This paper demonstrates the quantization of a spatial Cournot
duopoly model with product choice, a two stage game focusing on
non-cooperation in locations and quantities. With quantization, the players can access a continuous set of strategies, using continuous variable quantum
mechanical approach. The presence of quantum entanglement in the initial state identifies a quantity equilibrium for every location pair choice with any transport cost. Also higher profit is obtained by the firms at Nash equilibrium. Adoption of quantum strategies rewards us by the existence of a larger quantum strategic space at equilibrium.
\end{abstract}



\pacs{03.67.-a, 03.65.Ud,02.50.Le}

\maketitle

\section{Introduction}
Quantum entanglement, a physical resource, plays a vital  role in
quantum information theory: when used as a resource it performs
various tasks which seem rather impossible for classical resources.
In this regard quantum game theory is no exception, where the
concept of quantum entanglement has been  used for the benefit of quantum game over a classical one. Quantum game
theory  was first introduced by Meyer \cite{Mey99}, who showed that
a player can always beat his classical opponent by adopting quantum
strategies. Since then there has been a great deal of theoretical
efforts to extend the classical game theory into the quantum domain
which established the fact that quantum games are more efficient
than it's classical counterpart at least in terms of information
transfer \cite{EWL99,BH01,LDM02,HFM10}. Recently, successful accomplishment of the experimental realization
of quantum games \cite{PSWZ07,SFWKWH10,DLXSWZH02} on NMR quantum computer has
increased the attention on this topic.\\

In the literature of quantum games, one may note that the mostly
studied quantum games focus on games in which the players have
finite number of strategies. But in the economics of real life,
there are games in which the players can access to a continuous set
of strategies \cite{ball}, a classic example of which is the
Cournot's duoploy, a game which describes market competition. The
intimate connection between the game theory and the theory of
quantum communication motivated Li et al \cite{LDM02} to analyse the
quantization of Cournat's dupoly  using continuous variable quantum
system. Extending this idea we have studied the quantum equilibrium
for the Hotelling-Smithies model of product choice, which is a
spatial Cournot duopoly with transport cost and price
indiscrimination \cite{hotelling,Smi41}. Quantization of the game
shows that in equilibrium, the gain in the quantum domain is more,
when the transportation cost is low. For higher transport costs, the
difference between the quantum and classical profit is negligible.
We have also shown that if the consumers are supplied with a
$traveling$ $allowance$ to reach the seller, the optimal quantum
strategic space becomes larger than the classical optimum strategic
space.\\

In Section II we recapitulate the classical model of
Hotelling-Smithies with product choice. Section III is devoted to
the study the quantum version of the classical game and also  deals
with the benefit of the quantum game over the classical one. In
Section IV we treat the game with travelling allowance instead of
transport cost. We end with a conclusion.

\section{Classical Model of Hotelling-Smithies with product choice}

Hotelling \cite{hotelling} was the first to suggest that the
competition between oligopolistic sellers result in consumers being
offered products with an excessive sameness, where individual demand
is perfectly inelastic. His analysis was extended by Smithies
\cite{Smi41} to a case in which demand is inelastic and firms
compete in quantities. It has been recognised that Cournot
competition, which is quantity setting game, gives rise to different
equilibria than the Betrand model of competition which is price
setting game. We investigate the Hotelling-Smithies model of product
choice and study the Cournot equilibria for spatial duopolists which
are not able to price discriminate between their consumers. The
inability to price discriminate may arise for two different
situations: if consumers travel to the seller to collect the goods,
or if the seller is unable to customize the product to the
individual consumer's desires. \\

Let us first describe our model where  two firms indexed 1 and 2,
are assumed to choose locations on a linear market normalized to
unit length. Their locations, as measured by their distances from
the left side market boundary, are denoted by $r_1$ and $r_2$
respectively. From the symmetry we can assume $0\le r_1\le r_2\le
1$. According to the strategy of this two-stage game, these two
firms can sell a product that is homogeneous in all characteristics
other than the location at which it is available. Production cost is
normalized to zero for both of them and the firms are able and
willing to supply the entire market. They also simultaneously decide
that the firm i(i=1,2) will produce the product of quantity $q_i$
and sell them with price $p_i$ (per unit of product). We also
consider that the consumers are distributed uniformly with unit
density over the entire linear market. Consumer inverse demand
function is linear and is given by q(s)=1-p(s), where p is the
price, q is the supply and s denote the location of the consumer.
Inverse demand function and the positivity of q(s) and p(s) demand
that $q(s),p(s)\leq 1$. If t(assumed linear in distance and quantity
transported) is the transportation cost per unit length then the
consumer need to pay the price $p_i+t|s-r_i|$ for per unit product,
if he want to purchase the good from the i-th firm. It is very
natural that every consumer intends  to purchase the good with lower
price.\\

To solve the competitive game in the market we confine our
attention to a two stage game \cite{Smi41,chap7}. The firms first
choose locations and then quantities where the second stage
identifies the optimal quantity for any pair of locations of the
firms. In the first stage the equilibrium locations are derived in
the belief that the second stage choices will be an equilibrium
quantity pair for the second stage subgame. A pure strategy subgame
perfect Nash equlibrium for the two-stage quantity location Cournot
game is defined as a pair of locations $(r_1^c,r_2^c)$ and a
quantity pair $(q_1^c(r_1^c,r_2^c),q_2^c(r_1^c,r_2^c))$ such that

{\begin{equation}\label{qq1} \Pi_1((q_1^c(r_1^c,r_2^c),q_2^c(r_1^c,r_2^c)),
r_1^c,r_2^c)\geq \Pi_1((q_1^c(r_1,r_2^c),q_2^c(r_1,r_2^c)),
r_1,r_2^c) ~~~\forall~ r_1\in [0,1] \end{equation} and
\begin{equation} \label{qq2}\Pi_2((q_1^c(r_1^c,r_2^c),q_2^c(r_1^c,r_2^c)),
r_1^c,r_2^c)\geq \Pi_2((q_1^c(r_1^c,r_2),q_2^c(r_1,r_2)), r_1,r_2)
~~~\forall ~r_2\in [0,1] \ee} where {\be \Pi_1(
(q_1^c(r_1,r_2),q_2^c(r_1^c,r_2^c)), r_1^c,r_2^c)\geq
\Pi_1(q_1,q_2^c(r_1,r_2),r_1,r_2)~~ \forall ~q_1 \geq 0~
\rm{and}~r_1,r_2\in[0,1]\ee and \be \Pi_2(
(q_1^c(r_1,r_2),q_2^c(r_1,r_2)), r_1,r_2)\geq
\Pi_2(q_1^c(r_1,r_2),q_2^c(r_1,r_2)), r_1,r_2~~ \forall ~q_2
\geq 0~ \rm{and}~r_2\in[0,1]\end{equation}}

 The quantity subgame (\ref{qq1}) and (\ref{qq2}) can be solved for
two different cases: perfect agglomeration where $r_1=r_2$, which is
a symmetric case  and a general case where there is no perfect
agglomeration. In our model, we consider the case of non
agglomeration and also choose  $r_1 < r_2$ with $r_1\leq 0.5$ and
$r_2 \geq 0.5$. The firms are not agglomerated, the products are
perceived by consumers as being differentiated by location. We do
not consider the market overlap, given the outputs $q_i$ of the two
firms, the market clearing condition will determine the mill prices
$p_i$ (i=1,2). If $r$ is the market boundary\footnote{If $s\leq r$
consumer will purchase the good from firm 1 otherwise he will
purchase from firm 2.} between two firms then, \bea
p_1+t|r-r_1|&=&p_2+t|r-r_2|\nonumber\\
\mbox{or, ~~~~~ }r&=&\frac{p_2-p_1}{2t}+\frac{r_1+r_2}{2} \eea The
quantity $dq_1$ sold to consumers in the interval $ds$ located at
$s(0\le s \le r)$ is $dq_1=(1-p_1-t|s-r_1|)ds$ and the quantity
$dq_2$ sold to consumers in the interval $ds$ located at $s(r\le s
\le 1)$ is $dq_2=(1-p_2-t|s-r_2|)ds$ from which the aggregate
quantity sold by each firm is given by

\bea
\label{q1}q_1&=& \int_{0}^{r}dq_1\nonumber\\
&=&(1-p_1+tr_1)r-\frac{t(r)^2}{2}-tr_1^2\\
\label{q2}q_2&=&\int_{r}^{1}dq_2\nonumber\\
&=&(1-p_2+t(1-r_2))(1-r)-\frac{t(1-r)^2}{2}-t(1-r_2)^2 \eea
 The profit of  the firm $i$ is given by  \be \label{profitC}\Pi^C_{i}=p_iq_i ~(i=1,2) \ee
and  the Cournot reaction functions are {\bea
CR_1:&& \frac{\partial \Pi^C_{1}}{\partial q_1}=p_1+ q_1 \frac{\partial p_1}{\partial q_1}=0 \nonumber\\
\Rightarrow && 2tDp_1-[1-p_2+(1-r_2)+(1-r)][(1-p_1+tr_1)r-\frac{t(r)^2}{2}-tr_1^2]=0
\eea
\bea
CR_2:&& \frac{\partial \Pi^C_{2}}{\partial q_2}=p_2+ q_2 \frac{\partial p_2}{\partial q_2}=0 \nonumber\\
\Rightarrow &&
2tDp_2-[1-p_1+tr_1+tr][(1-p_2+t(1-r_2))(1-r)-\frac{t(1-r)^2}{2}-t(1-r_2)^2]=0
\eea} Here, $D=\frac{\partial F}{\partial p_1}\frac{\partial
G}{\partial p_2}-\frac{\partial F}{\partial p_2} \frac{\partial
G}{\partial p_1}$ and the demand functions can be written as the
implicit functions, \be
F(q_1,q_2,p_1,p_2)=0 \ee \be G(q_1,q_2,p_1,p_2)=0 \ee

The solutions of the  reaction functions give us the  price
$p_i(i=1,2)$ as functions of the location pair $(r_1,r_2)$ and the
solutions put into equations (\ref{q1}) and (\ref{q2}) give the Nash
equilibrium outputs $q^{C}_i(r_1,r_2)$ for the quantity subgame. The
analytical solutions for the reaction curves are difficult for its
complicated nonlinear nature. Equilibrium points for different sets
of location points $r_1$ and $r_2$ of the game can be solved
numerically for various transport costs. The detailed results are
found in \cite{chap7}. Computational results  indicate that for any
location pair $(r_1,r_2)$ an increase in the transport rate $t$
increase the mill prices, but reduces the quantity produced by each
firm and also reduce profits. In the classical game when the
duopolists are perfectly agglomerated i.e. when $r_1=r_2=r$, then
the output, price and profit will be greater the nearer are the
firms located to the market centre. This is a special case of the
standard Cournot model. We can summarize the result \cite{chap7} for
the classical quantity equilibrium  game as:\\
 For any location pair $(r_1,r_2)$ if firm i is located nearer the
 market centre than firm k, then
 \begin{itemize}
 \item firm i produces a greater output than firm k:\\
$q_1^c (r_1,r_2) \gtreqless q_2^c (r_1,r_2)$ if $r_1  \gtreqless 1-r_2$;
\item firm i will charge a higher mill price than firm k:\\
$p_1^c (r_1,r_2) \gtreqless  p_2^c (r_1,r_2)$ if $r_1  \gtreqless
1-r_2$; and
\item firm i will earn greater profit than firm k:\\
$\Pi_1^c (r_1,r_2) \gtreqless  \Pi_2^c (r_1,r_2)$ if $r_1
\gtreqless  1-r_2$
\end{itemize}

But, the above analysis does not indicate that a firm will always
wish to locate nearer to the market centre than its rival for
greater profit. Numerical results show that this will be the case,
if $t <t_g\cong 0.5104$ for any location pair $(r_1,r_2)$ with
$r_1\leq 0.5 \leq r_2$. When both the firms are located  very close
to the market centre i.e., $r_1,r_2 \rightarrow 0.5$ then,
$q_i^c(r_1,r_2)$ and $p_i^c(r_1,r_2)$ are,

\begin{eqnarray}
\lim \limits_{r_1,r_2 \to 0.5}q_i^c(r_1,r_2)&=&\frac{8-13t+\sqrt{97t^2+80t+64}}{48}\\
\lim \limits_{r_1,r_2 \to 0.5}p_i^c(r_1,r_2)&=&\frac{16+7t-\sqrt{97t^2+80t+64}}{24}
\end{eqnarray}

In the later part of our analysis it is shown that  the result
(Tables III and table IV) for the classical game with transport cost
$t=0.2$ and $t=0.6$, can be reproduced when the quantization
parameter is turned out to be zero.
\section{Quantum version of the Cournot Competition }
We now try to explore  the above discussed  classical game in
quantum domain by adopting the methodology described in
\cite{EWL99,LDM02,LK03}. We utilize a continuous set of eigenstates of
two single-mode electromagnetic fields which are initially in the vacuum state as \be \ket{\psi_0}=\ket{0}_1\otimes \ket{0}_2,
\ee where $\ket{0}_i(i=1,2)$ is the single-mode vacuum state
associated with the $i$-th electromagnetic field.
This state
consequently undergoes a unitary entanglement operation
$\hat{J}(\gamma)=\exp\{-\gamma(\hat{a_1}^{\dag}\hat{a_2}^{\dag}-\hat{a_1}\hat{a_2})\}$, where
$\hat{a_i}^{\dagger}(\hat{a_i})$ is the creation (annihilation) operator of the $i$'th
mode electromagnetic field. $\hat{J}(\gamma)$ is known to both firms and symmetric with respect to the
interchange of the two electromagnetic fields. Hence, the resultant state is given by
\be
\ket{\psi_i}= \hat{J}(\gamma)\ket{0}_1 \otimes \ket{0}_2
\ee
 The squeezing parameter $\gamma \geq 0$ is known as entanglement parameter and in the infinite squeezing limit
$\gamma \rightarrow \infty$, the initial state approximates the bipartite maximally entangled state
\cite{BK98,Fur98,MB99}. The strategic moves of firm 1 and firm 2 are represented by the
displacement operators $\hat{D_1}(x_1)$ and $\hat{D_2}(x_2)$ locally
acted on their individual fields. The players are restricted to
choose their strategies from the sets \beaa
S_j=\{\hat{D_j}(x_j)=exp(-ix_j\frac{i(\hat{a_j}^{\dag}-\hat{a_j})}{\sqrt{2}}|x_j\in[0, \infty)\}, \eeaa
where, $j=1,2$. After execution of their moves, firm
1 and firm 2 forward their electromagnetic fields to the final measurement, prior to which a disentanglement operation
$\hat{J}(\gamma)^{\dag}$ is carried out. Therefore the
final state prior to the measurement is \be \ket{\psi_f} =
\hat{J}(\gamma)^{\dagger}(\hat{D_1}(x_1)\otimes
\hat{D_2}(x_2))\hat{J}(\gamma)\ket{0}_1\otimes \ket{0}_2. \ee
One can set the final measurement such that it corresponds to the
 observables \beaa \hat{X_1} =
\frac{(\hat{a_1}^{\dag}+\hat{a_1})}{\sqrt{2}} & &\hat{X_2} =
\frac{(\hat{a_2}^{\dag}+\hat{a_2})}{\sqrt{2}}. \eeaa This
measurement is usually done by the homodyne measurement with assuming the state is infinitely squeezed.

The quantum game turns back to the original classical game when the
quantum entanglement is not present i.e.,
 $\gamma=0$. For, $\gamma=0$ the final measurement gives the
original classical results $q_1 = \langle \psi_f|\hat{X_1}|\psi_f\rangle=x_1$ and $q_2 = \langle
\psi_f|\hat{X_1}|\psi_f\rangle= x_2$.

On the other hand, for non-vanishing $\gamma$,
the final measurement gives the respective quantities of the two firms
\beaa
q_1 &=& x_1\cosh{\gamma}+x_2\sinh{\gamma}\\
q_2 &=& x_2 \cosh{\gamma}+x_1\sinh{\gamma} \eeaa
Referring to Eq.(\ref{profitC}), we obtain the  quantum profits for the firms $1$ and $2$
as \be\label{profitQ1}
\Pi^Q_1(x_1,x_2,p_1,p_2)=(x_1\cosh{\gamma}+x_2\sinh{\gamma})p_1 \ee
and  \be \label{profitQ2}
\Pi^Q_2(x_1,x_2,p_1,p_2)=(x_2\cosh{\gamma}+x_1\sinh{\gamma})p_2 \ee

Using the profit functions  the Cournot reaction
functions can be derived as \be CR_i: \frac{\partial \Pi_i}{\partial q_i}= p_i +q_i
\frac{\partial p_i}{\partial q_i}=0 \ee

Explicitly, the Cournot reaction functions for the two firms are respectively given by,\\

$CR^Q_1:$\\
{\begin{eqnarray}
-p_1 \cosh{\gamma}[(1-p_1+tr_1)(1-r)+(1-p_2+t(1-r_2))r]&&\nonumber\\
+(x_1\cosh{\gamma}+x_2\sinh{\gamma})[\cosh{\gamma}(1-p_2&&\nonumber\\+t(1-r_2)+t(1-r))+\sinh{\gamma}(1-p_1+tr_1-tr)]&=&0\label{cq1}
\end{eqnarray}}

$CR^Q_2:$\\
{
\begin{eqnarray}
-p_2 \cosh{\gamma}[(1-p_1+tr_1)(1-r)+(1-p_2+t(1-r_2))r]&&\nonumber\\
+(x_2\cosh{\gamma}+x_1\sinh{\gamma})[\cosh{\gamma}(1-p_1+tr_1)+tr)&&\nonumber\\+\sinh{\gamma}(1-p_2+t(1-r_2)-t(1-r))]&=&0\label{cq2}
\end{eqnarray}}
These quantum reaction functions can be solved for price
$p_i(i=1,2)$ as functions of the location pair $(r_1,r_2)$ and the
solutions put into equations (\ref{q1}) and (\ref{q2})  give the
Nash equilibrium outputs $q^{Q}_i(r_1,r_2)$ for the quantity
subgame.
\begin{figure}
\begin{center}
\includegraphics{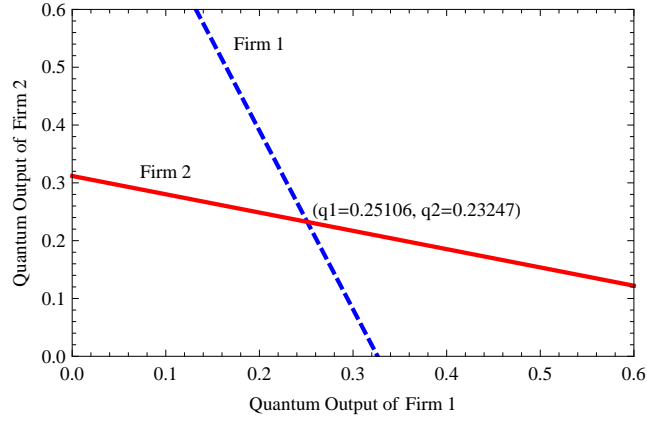}
\caption{\label{QReactionCurve} (Color online) The quantity reaction
curve for $\gamma=5, r_1=0.45, r_2=0.75, t=0.2$}
\end{center}
\vspace{-0.5cm}
\end{figure}

\begin{figure}
\begin{center}
\includegraphics{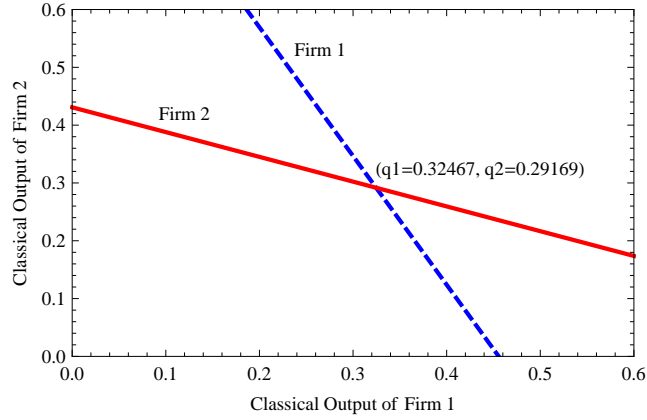}
\caption{\label{CReactionCurve} (Color online) The quantity reaction
curve for $\gamma=0, r_1=0.45, r_2=0.75, t=0.2$}
\end{center}
\vspace{-0.5cm}
\end{figure}
Figure \ref{QReactionCurve} shows the equilibrium point for the Firm 1 and Firm 2 in
the quantum game for the entanglement operator $\gamma=5$. As
expected the classical game is found to be a subset of this quantum
structure and our result for $\gamma=0$, shown in Fig. \ref{CReactionCurve}, reproduces
the classical result given in \cite{chap7}. Tables
\ref{fig:tableQT02g5} \& \ref{fig:tableQT06g5} provide more detailed
information on the quantum equilibria of
 this quantity game. For every table including Tables \ref{fig:tableQT02g5} \& \ref{fig:tableQT06g5}, the data given in the sub-tables are for firm 1. Transposing these sub-tables gives the corresponding data for firm 2. For example, for the location pair $(r^i_1,r^j_2)$ if the output of firm 1 is $q_{ij}$, then output of firm 2 is $q_{ji}$.

\begin{table}
\caption{\label{fig:tableQT02g5} Quantum solutions of the game for t=0.2, $\gamma=5$. Data in sub-tables are for firm 1. Transposing these sub-tables gives the corresponding data for firm 2. For example, for the location pair $(r^i_1,r^j_2)$ if the output of firm 1 is $q_{ij}$, then output of firm 2 is $q_{ji}$. }
\vspace{-1.4cm}\begin{center}
\scalebox{0.65}{
\includegraphics{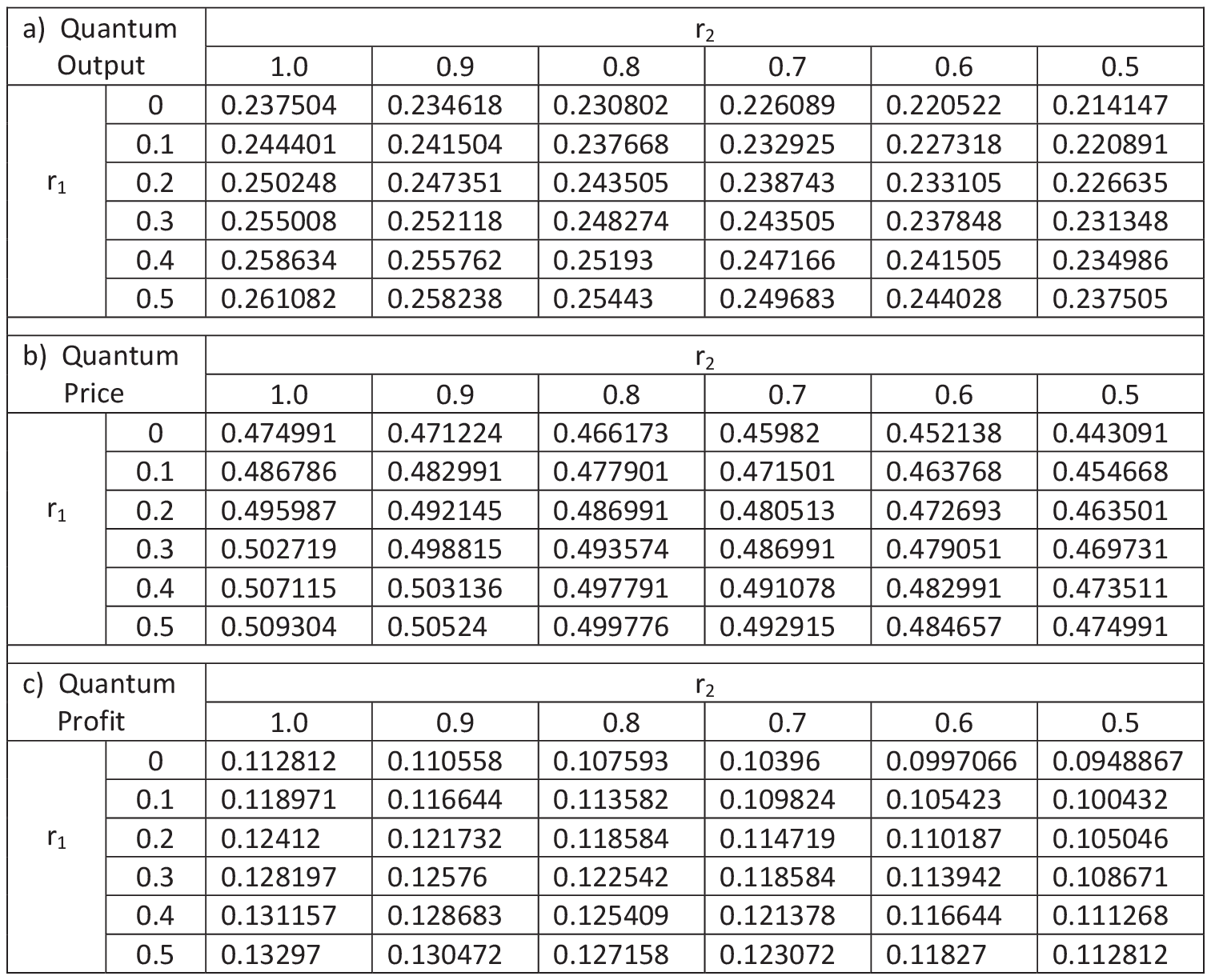}
}\end{center}\vspace{-9.6cm}
\end{table}

\begin{table}
\caption{\label{fig:tableQT06g5} Quantum solutions of the game for t=0.6, $\gamma=5$.}
\vspace{-1.1cm}
\begin{center}\scalebox{0.65}{
\includegraphics{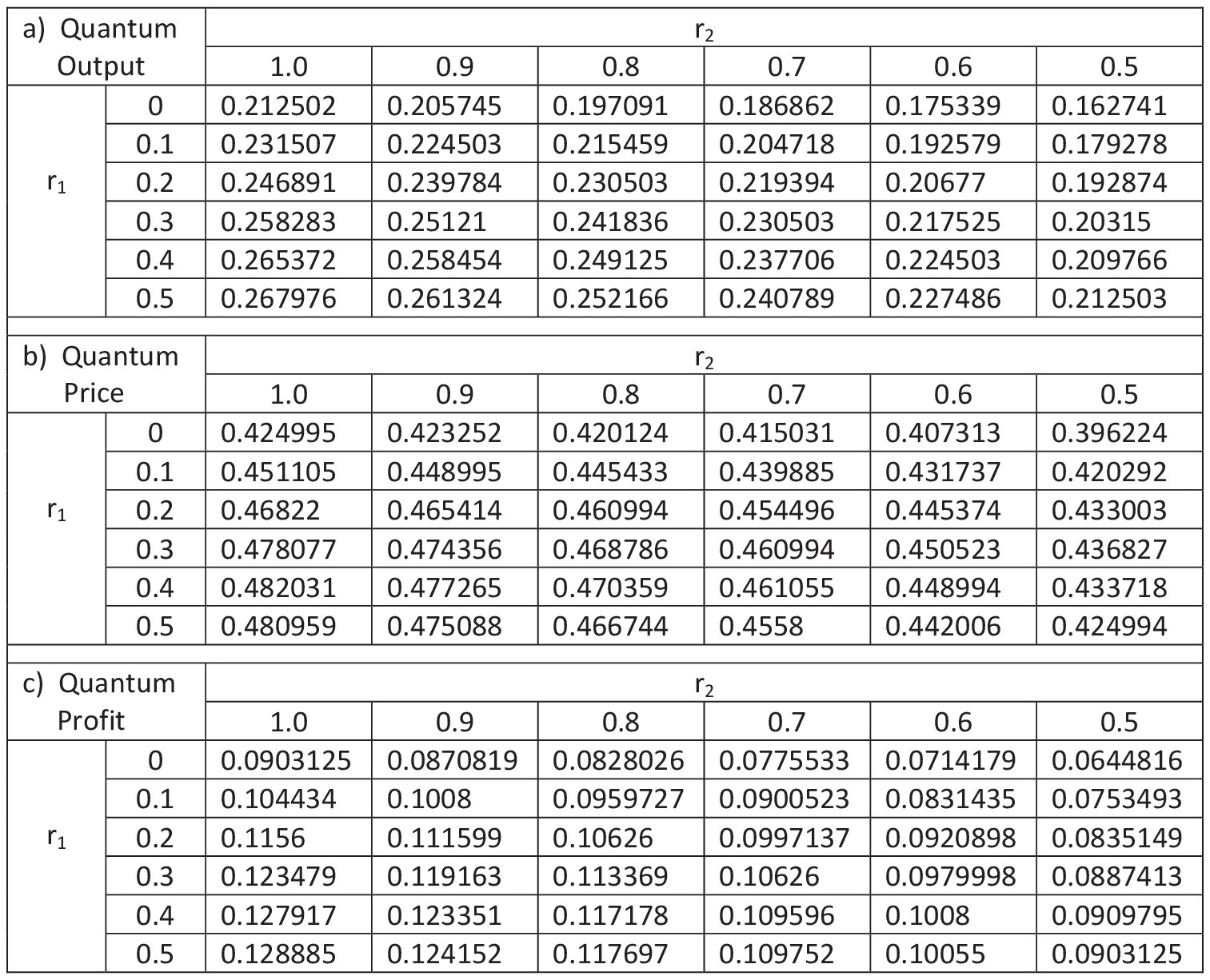}
}\end{center}
\vspace{-9.8cm}
\end{table}
Tables \ref{fig:tableT02g0} \& \ref{fig:tableT06g0} show that
setting $\gamma=0$, in our theory we can reproduce the classical
game results given in \cite{chap7}.

\begin{table}
\caption{\label{fig:tableT02g0}Classical(i.e. $\gamma=0$) solutions of the game for t=0.2.}
\vspace{-1.4cm}
\begin{center}\scalebox{0.65}{
\includegraphics{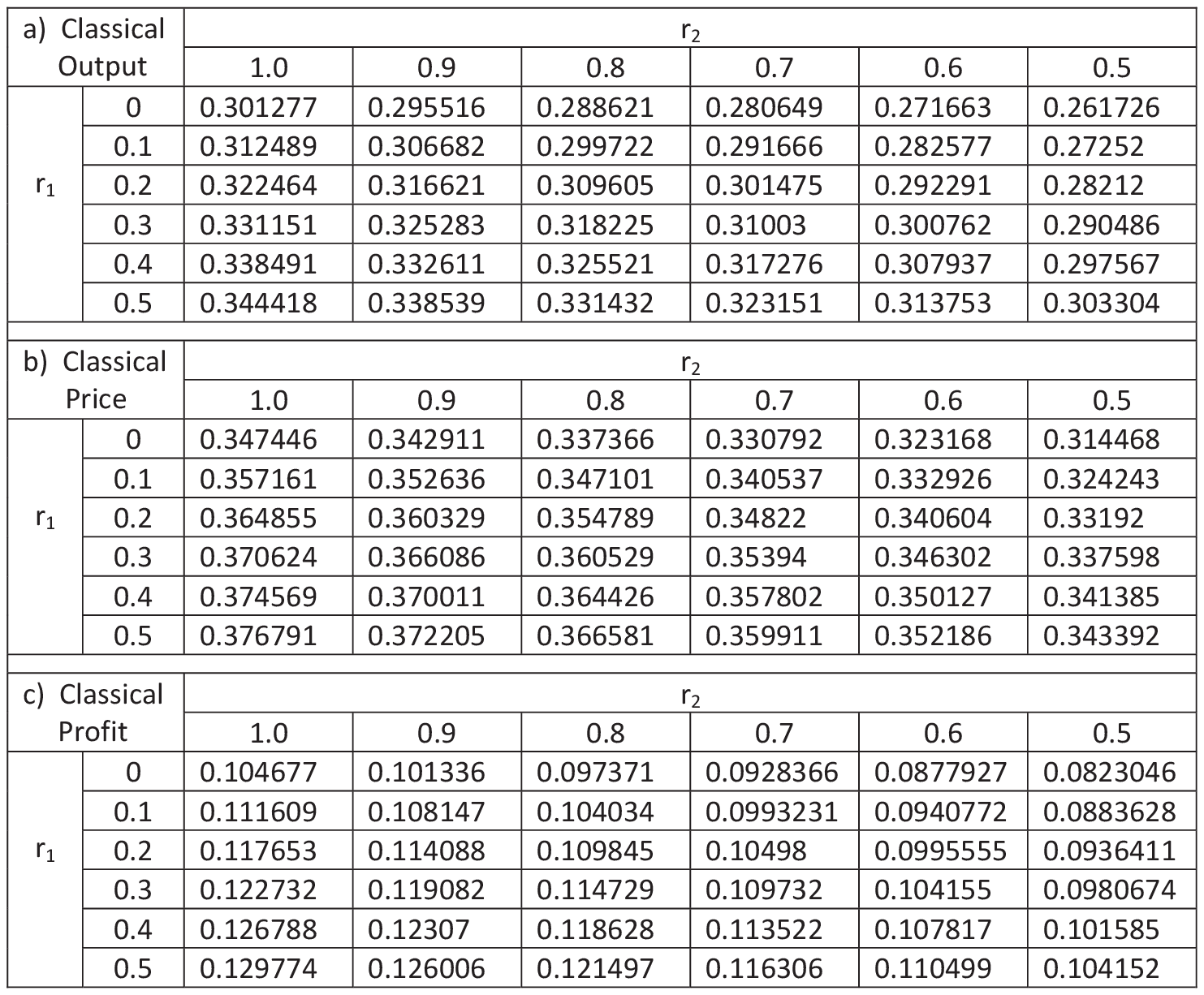}
}\end{center}
\vspace{-9.5cm}
\end{table}

\begin{table}
\caption{\label{fig:tableT06g0} Classical(i.e. $\gamma=0$) solutions of the game for t=0.6. }
\vspace{-1.4cm}
\begin{center}\scalebox{0.65}{
\includegraphics{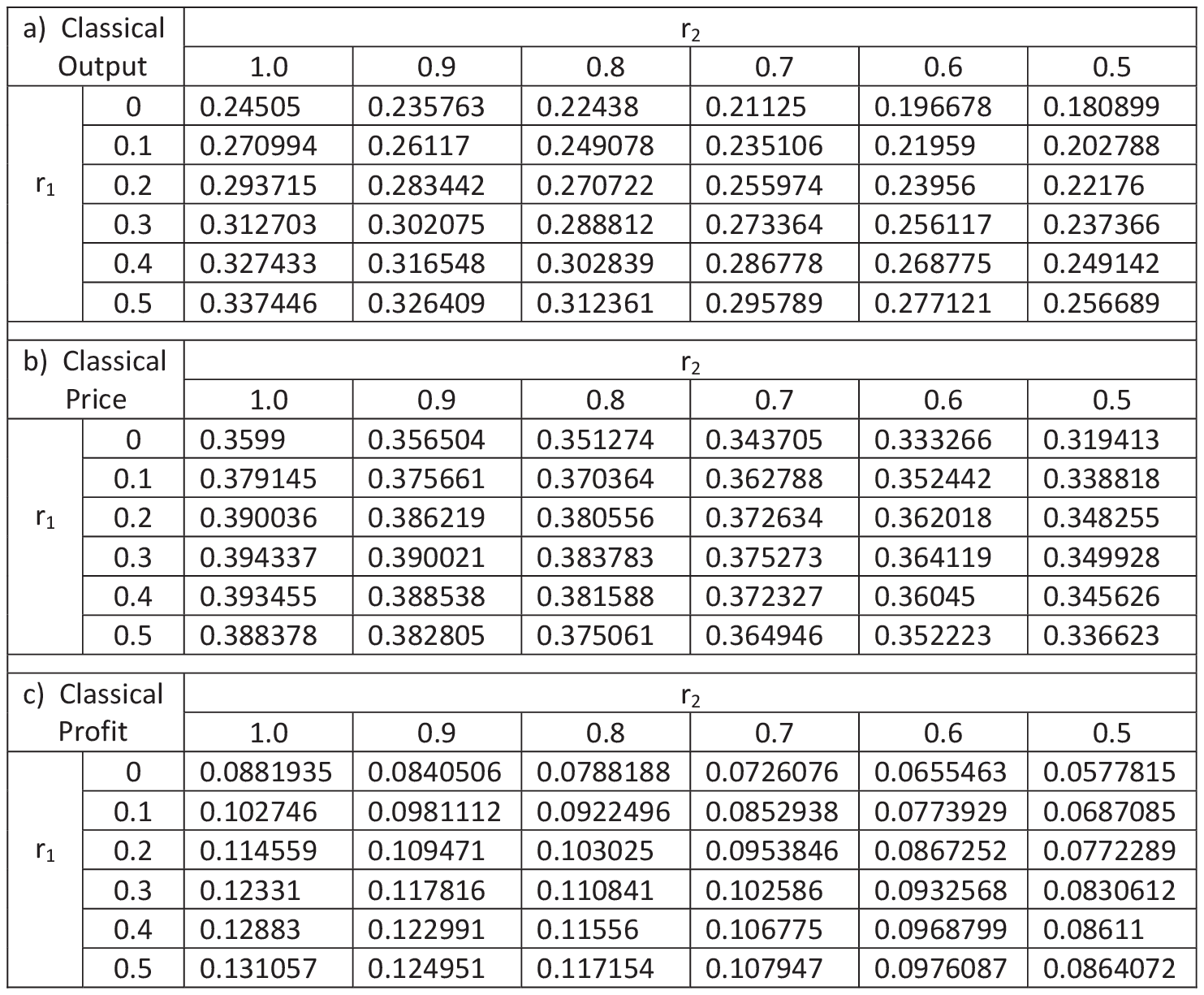}
}\end{center}\vspace{-9.6cm}
\end{table}

A careful examinations indicate that for any location pair
$(r_1,r_2),$ an increase in the transport rate $t$ increase the mill
prices, but reduces the quantity produced by each firm and also
reduce profits. The results also provide a benchmark against which
we can assess the actual locations the firms choose in the location
stage of the quantity location game. Independent of the transport
rate, when the firms are located symmetrically (i.e. $r_2=1-r_1)$,
the individual firm profit is maximised when the firms are located
inside but $near$ the quartiles. The aggregate output is maximized
when the firms are located symmetrically. But surprisingly, the
symmetric location pair that maximises aggregate output is more
agglomerated when the transport rate is higher. Low transport costs
encourage agglomeration and
 quantity competition. Higher transport costs imply that sales
 decline relatively quickly with distance from the firm and so give
 heavier weight to consumers close to the firm. This moderates
 somewhat the competitive pressures of proximate  locations.\\

\begin{figure}
\begin{center}\scalebox{0.9}{
\includegraphics{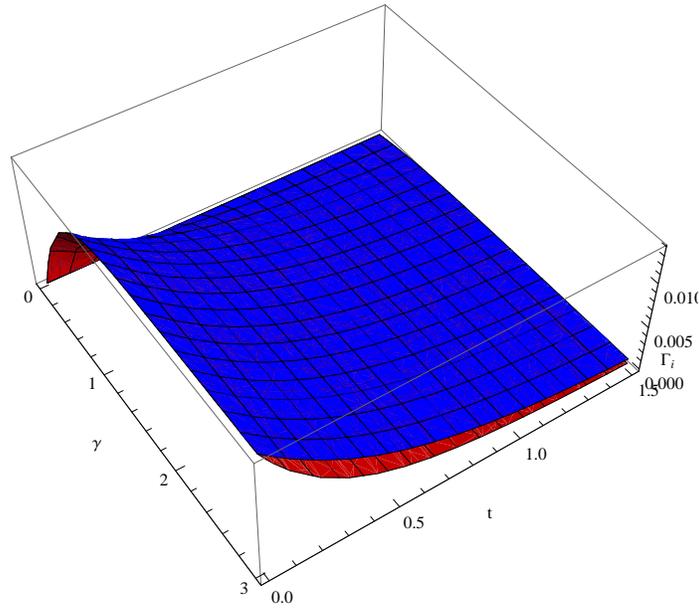}}
\caption{\label{fig:3DPG} (Color online) The quantum benefit $\Gamma_i=\Pi_i^Q-\Pi_i^C$ at equilibrium point over the classical equilibrium as a function of
the transportation cost `t' and the entanglement parameter `$\gamma$'. We have selected
the location of the firm 1 $r_1=0.3$ and the location of the firm 2
$r_2=0.6$. Blue plot: Profit for firm 1 w.r.t. transportation cost
`t' and `$\gamma$'. Red plot: Profit for firm 2 w.r.t.
transportation cost `t' and `$\gamma$'.}
\end{center}
\end{figure}

Another intriguing  result is expressed  in Fig. \ref{fig:3DPG}.
Figure \ref{fig:3DPG} needs some explanation. It is noted
that the quantum benefit of the profit of the firm 2, $\Gamma_2=\Pi_2^Q-\Pi_2^C$ is greater than
$\Gamma_1=\Pi_2^Q-\Pi_2^C$(quantum benefit of the profit of the firm 1) for higher transport cost, whereas it
decreases for lower $t$ and finally the difference
$\Gamma_2-\Gamma_1=0$ for $t=0$ and $\gamma=0$. $\Gamma_2 \geq \Gamma_1$, due to the fact that the location $r_2=0.6$ of the firm 2 is towards more central than the location $r_1=0.3$ of the firm 1. So there is a strong quantum advantage for firm i if the location of the i-th firm is more central. 

\begin{figure}
\begin{center}
\includegraphics{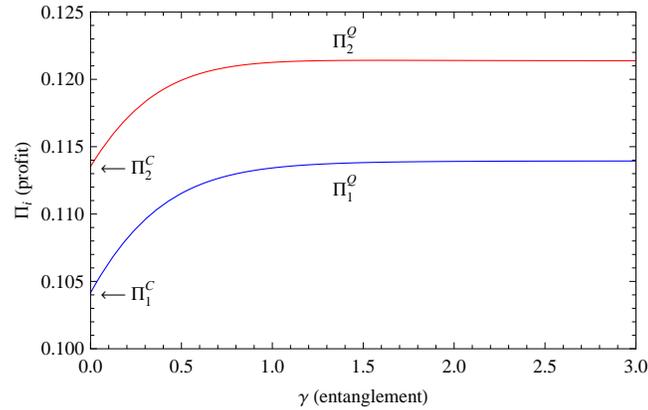}
\caption{\label{fig:t3x3x6} (Color online) Quantum profit at Nash
equilibrium w.r.t. the entanglement parameter $\gamma$ for t=0.2,
$r_1=0.3$ and $r_2=0.6$. Blue line: Quantum profit ($\Pi_1^Q$) of
the firm 1 w.r.t. $\gamma$. Red line: Quantum profit($\Pi_2^Q$) of
the firm 2 w.r.t. $\gamma$. Here $\Pi_i^C$ denoted the classical
profit of the i-th firm.}
\end{center}
\end{figure}

The plot of the profits for both the firms with the entanglement
parameter $\gamma$, for a fixed transport cost and a set of a fixed
location parameters given in Figure \ref{fig:t3x3x6} explains
vividly the quantum benefit over its classical counterpart.

\begin{figure}
\begin{center}
\includegraphics{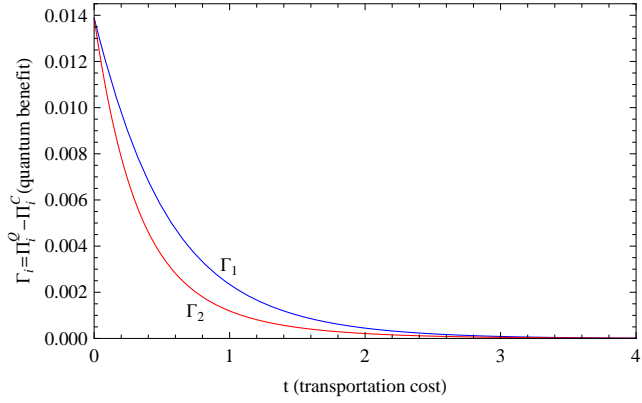}
\caption{\label{fig:profitgain} (Color online) The  quantum benefit $\Gamma_i=\Pi_i^Q-\Pi_i^C$ at Nash equilibrium point over the classical Nash equilibrium as a function of
the transportation cost `t'. Here the entanglement parameter
$\gamma=5$, location of the firm 1 $r_1=0.3$ and location of the
firm 2 $x_2=0.6$. Blue line: Quantum benefit($\Gamma_1$) for firm 1 w.r.t.
transportation cost `t'.
 Red line: Quantum benefit $\Gamma_2$ for firm 2 w.r.t. transportation cost `t'.}
\end{center}
\end{figure}

\begin{figure}
\begin{center}
\includegraphics{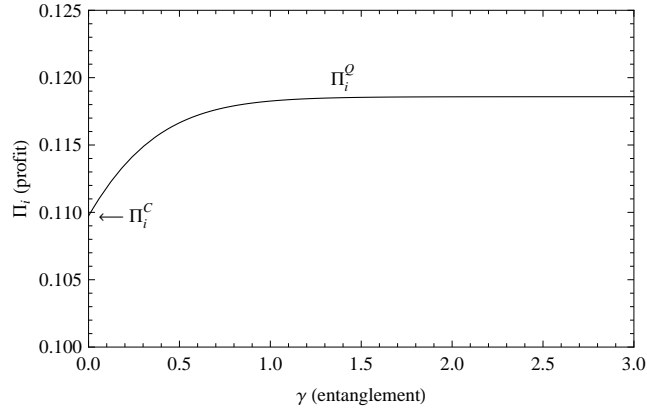}
\caption{\label{fig:symmetriclocation} The profits at quantum Nash equilibrium w.r.t the entanglement
parameter $\gamma$, when the two firms are perfectly agglomerated or, symmetric location(i.e. $r_1=1-r_2$) for t=0.2, $r_10.3$ and $r_2=0.7$. Here both the quantum (classical) profits are same, i.e. $\Pi_1^Q=\Pi_2^Q$ ($\Pi_1^C=\Pi_2^C$).}
\end{center}
\end{figure}

Next, one  can analyse the quantum benefit
$\Gamma_i=\Pi_i^Q-\Pi_i^C$ for a fixed $\gamma$ (say, $\gamma=5$)
and a fixed set of locations of the firms for different values of
the transport cost $t$. The analysis is shown in Fig.
\ref{fig:profitgain}. The quantum benefit is
maximum for zero transport cost for both the firms. 
 As $t$ is increased the benefit rapidly falls for both the firms,
but the rate of decrement is more for firm 2 than firm 1 (as we discussed earlier this is also due to the fact that the location of firm 1 is more central also seen in the Fig. \ref{fig:3DPG}), and as expected when $t$
attains a much higher value,  the quantum benefit $\Gamma_i$ asymptotically
tends to zero for both the firms, i.e.
 for a higher transport cost the quantum profit over its classical counterpart is
 negligible.
The existence of a symmetric Nash equilibrium (i.e., profit of firm
1= profit of firm 2) for the zero consumer transportation cost is an
example of another important feature of this game. We can observe
the same feature if $t$ is independent of the distance. The
symmetric quantum Nash equilibrium is also obtained if the firms are
perfectly agglomerated ($r_1=r_2$) or symmetrically located
($r_1=1-r_2$).
  A key feature in Li. et. al \cite{LDM02} is the zero transportation cost or the firms location are perfectly
  agglomerated and the quantum profit gain at Nash equilibrium over classical is equivalent to the case given in
  figure \ref{fig:symmetriclocation}, where the locations of the firms are considered to be symmetric within the
  market i.e. $r_1=1-r_2$.\\

The quantization of a classical game can be termed as successful
when  the quantum profit is higher than the classical profit. A
critical comparison of the Tables \ref{fig:tableQT02g5},
\ref{fig:tableQT06g5}, \ref{fig:tableT02g0} and \ref{fig:tableT06g0}
very nicely explain the competency of our model in this regard by
showing  that  for different transport costs,
 at equilibrium points, the quantum profit is more than the corresponding classical
 profit. The behaviour in outputs of the firms for the quantum version of the game is similar
 to that of its classical counterpart, but with higher profit. Like classical situation, in quantum case there also exists a transport cost $t_g^Q(\gamma)$ ($\cong  0.39353;$ when, $\gamma=5$) for which, any $t<t_g^Q(\gamma)$ firm i's profit is always greater, if the location of the firm is nearer to the market centre. Also,

{\small \begin{eqnarray}
\lim \limits_{r_1,r_2 \to 0.5}q_i^Q(r_1,r_2)=[(8 - 13 t)\cos{\gamma}- t\sin{\gamma}+\surd((64 + t (80 + 97 t))\cos^2{\gamma} \nonumber\\+ t\sin{\gamma}(2(40 + t)\cos{\gamma}+ t\sin{\gamma}))]/(16(3\cos{\gamma}+\sin{\gamma}))\\
\nonumber\\
\lim \limits_{r_1,r_2 \to 0.5}p_i^Q(r_1,r_2)=[(16 + 7 t)\cos{\gamma} + \sin{\gamma}(8 - t)-\surd((64 + t (80 + 97 t))\cos^2{\gamma} \nonumber\\+ t\sin{\gamma}(2(40 + t)\cos{\gamma}+ t\sin{\gamma}))]/(8 (3\cos{\gamma} + \sin{\gamma}))
\end{eqnarray}}
\section{Game with transport allowance}
Finally, in this section  we describe a fascinating situation. Let
us think of the case when a consumer does not bear the transport
cost, instead the consumer earns some amount of money as  $transport$
$allowance$ for his/her transportation to the firm for buying any
product . This situation can be explored by setting $t$ a negative
value, where each consumer earns an amount $u=-t$ per unit distance
when they travel to the seller(Firm) to collect the goods. We call
it $transport$ $allowance$ and denote it as
  $u=-t$ per unit distance.

Numerical computations  are displayed in the  Tables \ref{fig:tableCnT02g0} \& \ref{fig:tableQnT02g5}.\\
 \begin{table}
\caption{\label{fig:tableCnT02g0} Classical(i.e., $\gamma=0$) solutions of the game for t=-0.2.}
\vspace{-1.2cm}
\begin{center}\scalebox{0.65}{
\includegraphics{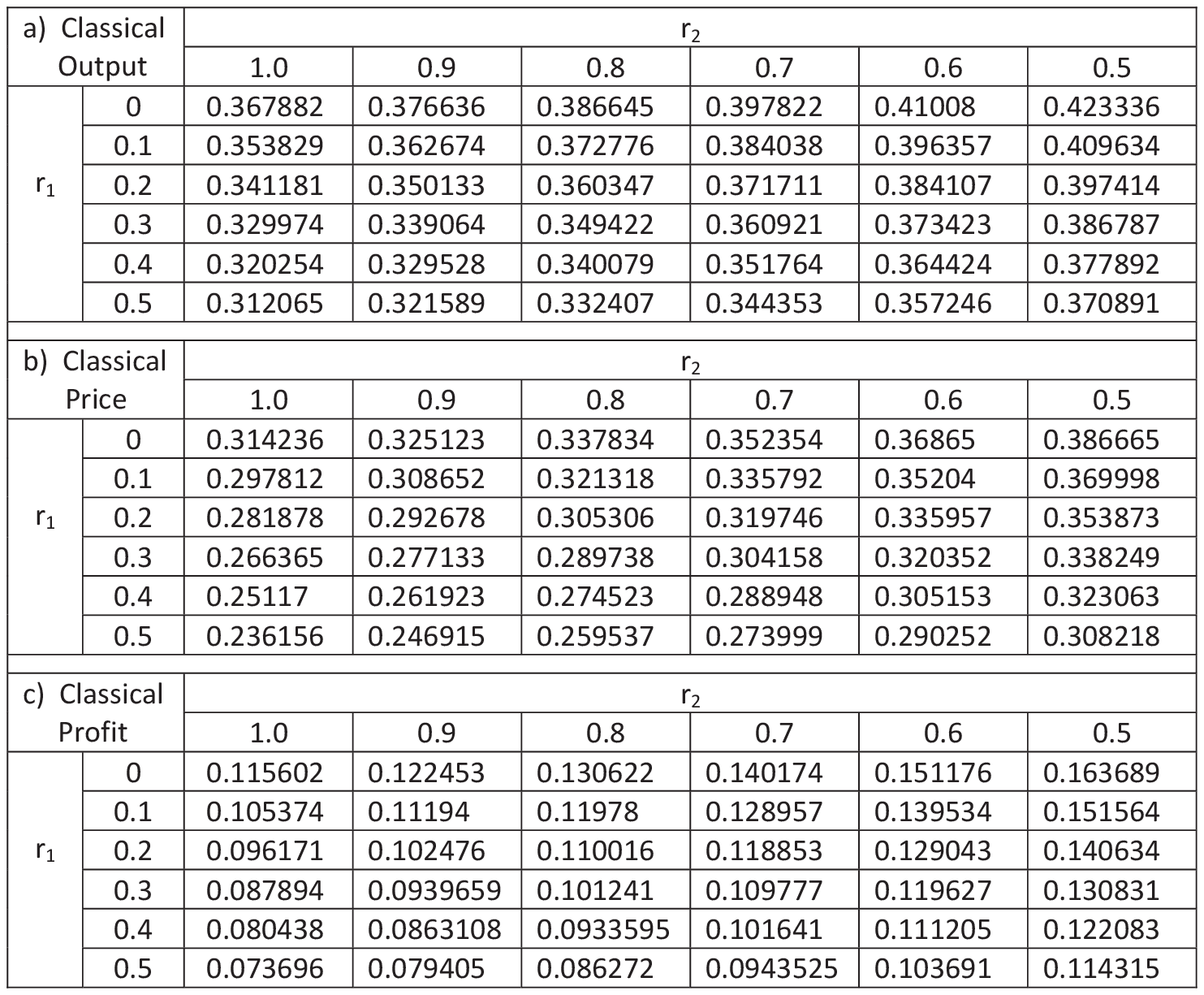}
}\end{center}
\vspace{-9.0cm}
\end{table}

  \begin{table}
  \caption{\label{fig:tableQnT02g5} Quantum solutions of the game for t=-0.2, $\gamma=5$. }
\vspace{-1.2cm}\begin{center}
\scalebox{0.65}{\includegraphics{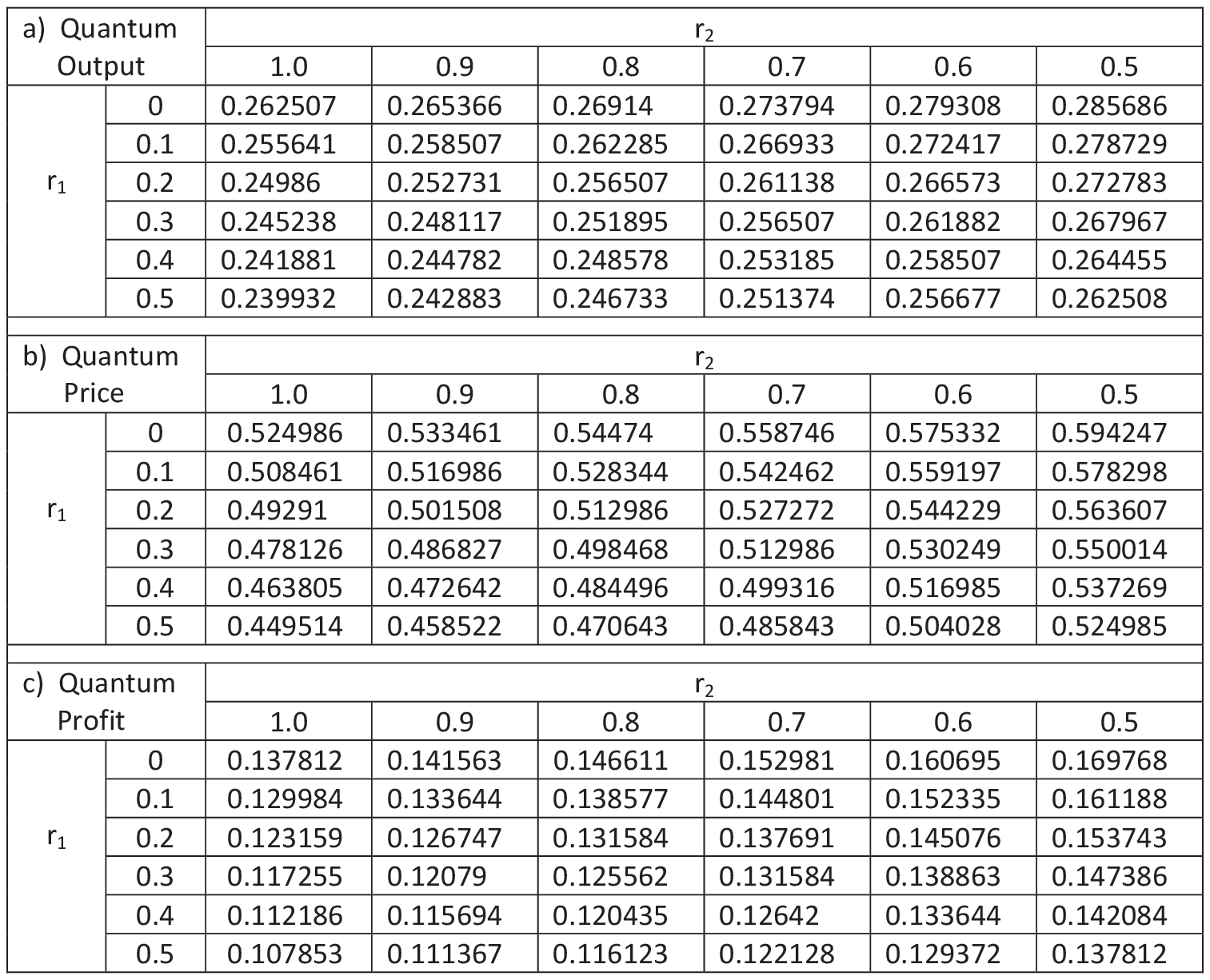}
}\end{center}\vspace{-9.0cm}
\end{table}

\begin{figure}
\begin{center}
\scalebox{0.75}{\includegraphics{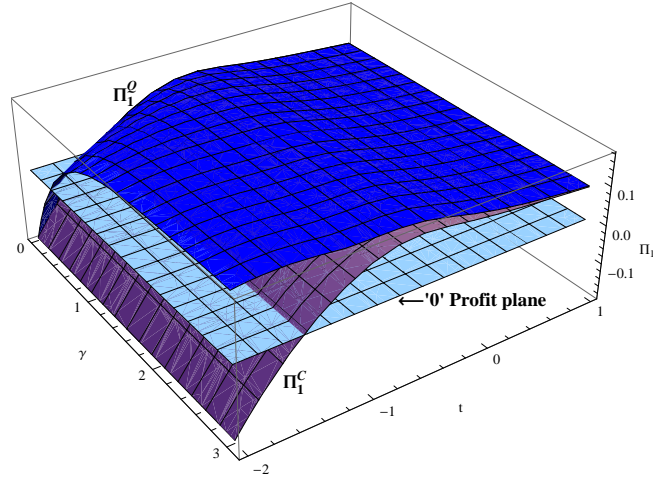}}
\caption{\label{fig:Negative_t1} (Color online) Plot of quantum
(blue surface) and classical (pink surface) profits with respect to
$\gamma$ and t for the location $r_1=0.3$ and $r_2=0.6$ of firm 1 at
equilibrium point. Plot shows that for a large region of $\gamma$
and $negative$ `t', the quantum profit is still above(i.e. positive) the
`0' profit plane (cyan surface) , whereas classical profit is
negative (i.e. below the `0' profit plane).}
\end{center}
\end{figure}

\begin{figure}
\begin{center}
\scalebox{0.75}{\includegraphics{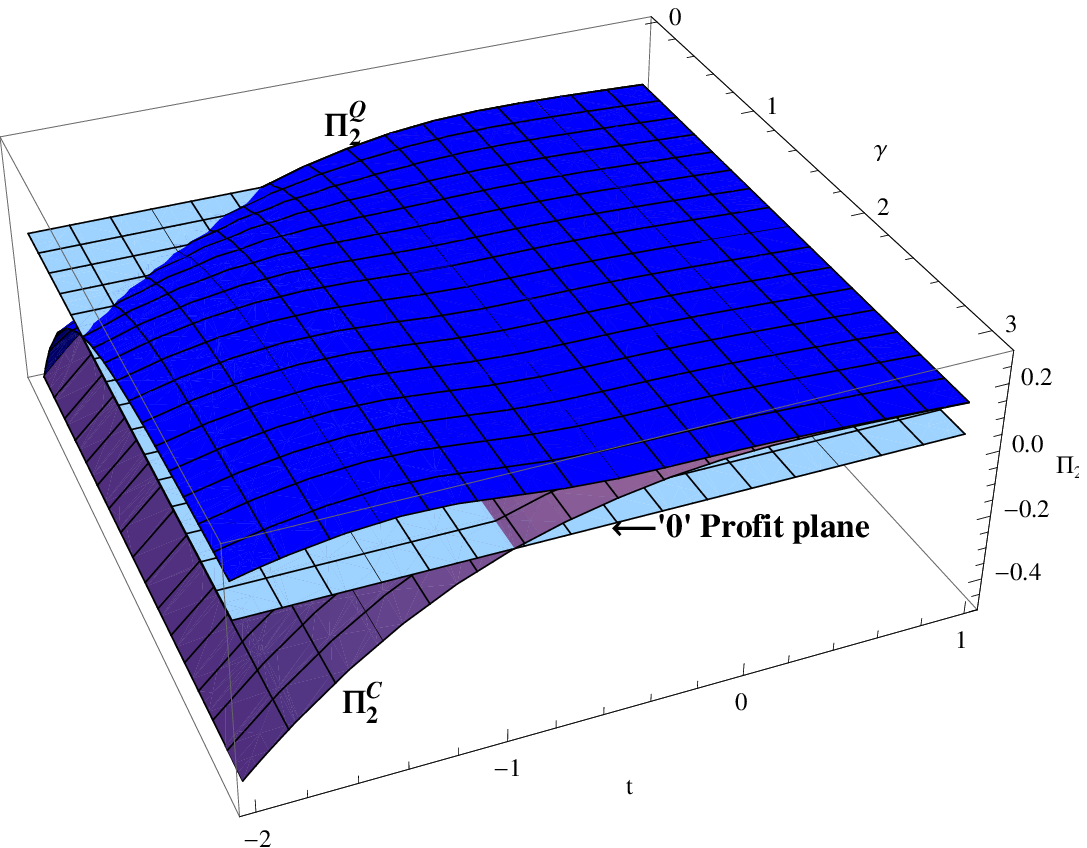}}
\caption{\label{fig:Negative_t2} (Color online) Plot of quantum
(blue surface) and classical (pink surface) profits with respect to
$\gamma$ and t for the location $r_1=0.3$ and $r_2=0.6$ of firm 2 at
equilibrium point. Plot shows that for a large region of $\gamma$
and $negative$ `t', the quantum profit is still above(i.e. positive) the
`0' profit plane (cyan surface) , whereas classical profit is
negative (i.e. below the `0' profit plane).}
\end{center}
\end{figure}

Next, we analyse the variation of the profit with the transport allowance $-t=u$ and entanglement parameter $\gamma$.
Figures  (\ref{fig:Negative_t1}\& \ref{fig:Negative_t2}) show the
variation of the profit $\Pi_i$ (i=1,2) for a wide range of t,
($-1.5<t<1$) and $\gamma$.
We denote $t=t_c$ as a critical point for
the profit of the firm $i$, $\Pi_i$, when the profit is zero and after that starts to be negative ( $\Pi_i <0$). For any $\Pi_i<0~(i=1,2)$, it is seen that for  $t<t_c<0$, at least one optimal profit is negative. We can write $-t_c=u_c$ as the critical $travel$
$allowance$.

The $t$-$\gamma$ plane with $\Pi_i=0$ is denoted as the zero profit
plane. It is noticed that for a large region of $\gamma$ and $-ve$
`t', the quantum profit is still above the `0' profit plane (i.e.
positive), whereas in the classical case, profit is  below the `0'
profit plane which means profit is negative.

With this analysis,  another important feature of quantum game theory can be explored.
Let $SS=\{(q_1,q_2,p_1,p_2)|0\leq q_i, p_i (i=1,2)\}$ be the
Strategic Space(SS) and $OSS$ be the Optimal Strategic
Space(OSS)\footnote{set of strategic points where equilibrium take
place.} of a quantity equilibrium game for fixed location.
Therefore, $OSS\subset SS$. \\
For $\Pi_i=q_ip_i<0 \Rightarrow$ we should have either $p_i<0$ or, $q_i<0$, i.e., for negative profit the equilibrium
strategic
  point $(q_1,q_2,p_1,p_2) \notin SS (\supset OSS)$.\\

Figures (\ref{fig:Negative_t1}\& \ref{fig:Negative_t2}) show that for both the firms $t_c^Q<t_c^C<0$ for a fixed location ($r_1 \& r_2$), here $-t_c^Q$ is the genuine($\gamma >0$) quantum critical $travel$ $allowance$ $u_c^Q$, 
and $u_c^C=-t_c^C $ is the classical critical
  $transport$ $allowance$. Therefore, $\exists~~ t(< t_c^C<0)$, such that both the quantum profits $\Pi_1^Q \& \Pi_2^Q >0$,
  whereas at least one classical profit (either $\Pi_1$ or, $\Pi_2$) is negative, for a fixed location.  So the optimal
  quantum strategic space($OSS^Q$) is always larger than the optimal classical strategic space ($OSS^C$),
  which is another feature of importance in a quantum game.

The plots (\ref{fig:Negative_t1} \& \ref{fig:Negative_t2}) and the
tables (\ref{fig:tableCnT02g0} \& \ref{fig:tableQnT02g5}) indicate
that if the present two-stage game is played, either classically or
in the quantum domain, with transport  allowance instead  of
transportation  cost
 then: \\
 For any location pair $(r_1,r_2)$ if firm 1 is located nearer the
 market centre than firm 2, then\\
 \begin{itemize}
 \item firm i produces a greater output than firm k:\\
$q_1^c (r_1,r_2)  \gtreqless  q_2^c (r_1,r_2)$ if $r_1  \lesseqgtr
1-r_2$;
\item firm i will charge a higher mill price than firm k:\\
$p_1^c (r_1,r_2)  \gtreqless  p_2^c (r_1,r_2)$ if $r_1  \lesseqgtr
1-r_2$; and
\item firm i will earn greater profit than firm k:\\
$\Pi_1^c (r_1,r_2) \lesseqgtr  \Pi_2^c (r_1,r_2)$ if $r_1
\gtreqless  1-r_2$
\end{itemize}
which implies that for negative `t', there is a strong competitive advantage when the
 firm location are nearer the end of the market boundary unlike the case when the transport cost `t' is positive.

\section{Conclusion}
In this Letter, we explore some interesting cases of Hotelling-Smithies model
 of product choice where the player can be benefited more by adopting a quantum strategy rather
the classical strategy, leaving a large number of cases unresolved.
Specially, the quantum benefit in Bertrand competition (with
transport cost indiscrimination) which seems a more difficult problem
than the case of Cournot quantity competition. We also demonstrate
the fact that the quantum equilibrium strategic space of a Cournot
quantity competition game is larger  than the classical equilibrium
strategic space. Although, this result as expected, is uncommon for
any quantum game having only linear demand function and  hope that
this result would encourage the researchers in this area to develop
the field further.

\section{acknowledgments}
The authors acknowledge helpful discussions with Dr. Kausik
Gangopadhyay. R.R. also acknowledges support from Norwegian Research
Council.








\end{document}